\documentclass[useAMS,usenatbib]{mn2e}

\usepackage{graphicx}
\usepackage{natbib}

\title[JHK light curves of RR Lyr]{The infrared JHK light curves of RR Lyr}
\author[Sollima et al.]{A. Sollima$^{1}$\thanks{E-mail:
antonio.sollima@oabo.inaf.it (AS)}, C. Cacciari$^{2}$, A. A. Arkharov$^{3}$, 
V. M. Larionov$^{3,4}$, D. L. Gorshanov$^{3}$, 
\newauthor
N. V. Efimova$^{3}$ and A. Piersimoni$^{5}$ \\
$^{1}$Dipartimento di Astronomia, Universit\`a di Bologna, via Ranzani 1,
Bologna, 40127-I, Italy \\
$^{2}$INAF Osservatorio Astronomico di Bologna, via Ranzani 1,
Bologna, 40127-I, Italy \\
$^{3}$Pulkovo Observatory, Russian Academy of Sciences, St. Petersburg, Russia \\
$^{4}$Sobolev Astronomical Institute, St. Petersburg State University, Russia\\
$^{5}$INAF Osservatorio Astronomico di Collurania, via M. Maggioni, 64100-I
Teramo, Italy}
\begin{document}

\date{Accepted 2007 August ??; Received 2007 August ??; in original form
2007 July ??}

\pagerange{\pageref{firstpage}--\pageref{lastpage}} \pubyear{2007}

\maketitle

\label{firstpage}

\begin{abstract}
We present infrared JHK time series photometry of the variable star RR Lyr, 
that allow us to construct the first complete and accurate infrared light curves for 
this star. 
The derived mean magnitudes are $<J>$=6.74 $\pm$ 0.02, 
$<H>$=6.60 $\pm$ 0.03 and $<K>$=6.50$\pm$ 0.02.
The $<$K$>$ magnitude is used to estimate the reddening, the mass, 
the mean luminosity and temperature of this variable star.
The use of these RR Lyr data provide a more accurate absolute calibration 
of the P-L$_K$-[Fe/H] relation, and a distance modulus 
$(m-M)_{0}=18.48\pm0.11$ to the globular cluster Reticulum in the LMC.      
\end{abstract}


\section{Introduction} \label{int}

RR Lyr is the nearest (i.e. brightest) and best studied variable star of its 
type and therefore plays a crucial role in setting the zero-point of 
the distance scale.  
There is a very extensive literature reporting photometric measurements and 
light curves
in all optical bands, as well as spectroscopic measurements and radial velocity 
curves (see e.g. Preston et al. 1965; Wilson et al. 1989; Szeidl et al. 1997,
and references therein). The pulsation period is known to a high degree of 
accuracy ($P=0.5668386\pm0.0000016$ d, Kolenberg et al. 2006) and it is well 
established that RR Lyr shows the Blazhko effect (i.e. a modulation of its
amplitude) with a period of 38.4 d (Kolenberg et al. 2006). The amplitude of
this modulation also varies over a 4-yr period (Detre \& Szeidl 1973). 

Thanks to its proximity, RR Lyr has the most accurately determined 
trigonometric parallax. In particular, Benedict et al. (2002) derived the
absolute parallax $\pi=3.82\pm0.20~mas$, corresponding to a distance 
modulus of $(m-M)_{0}=7.09\pm0.11$, from the analysis of HST data. 
A very recent reanalysis of Hipparcos data (van Leeuwen 2007) gives a 
value $\pi=3.46\pm0.64~mas$.  

However, infrared photometric observations are nearly missing for this star,
the only published data being the two data points (about half a cycle apart) 
by Fernley et al. (1993, hereafter FSB93) and the 2MASS single epoch data point. 
Based on these very scanty data, the mean K magnitude of RR Lyr has 
nevertheless been used to calibrate both theoretical and empirical period - K
luminosity relations (Bono et al. 2003; Sollima et al. 2006), but the need of 
a more reliable and accurate determination was strongly felt. 

This paper reports on the first infrared (JHK) complete light curves that we
have obtained for RR Lyr, derives a new estimate of reddening, distance and  
physical parameters for this star, provides a more reliable and accurate 
absolute calibration of the period - K luminosity relation and 
discusses the implications of these new results on the zero-point 
of the distance scale.   

\section{Observations and Data reduction} 
\label{reduct}

The observations have been made with the AZT24 1.1m telescope of the Campo 
Imperatore Observatory (L'Aquila, Italy), equipped with the NIR camera SWIRCAM 
(see Di Paola 2003, Brocato \& Dolci 2003 and Del Principe et al. 2005 for a 
detailed description of the telescope and the camera). The observations have
been taken on several nights during the period May-July 2007, in good
photometric conditions. The seeing (FWHM) ranged from 1.7 to 3.5 arcsec, with
the average value around 2.3 arcsec.  
We observed a 4.4x4.4 arcmin field around RR Lyr
through the J, H and K standard filters. Apart from RR Lyr, 7 fainter 
2MASS stars
are present in this field. Given the brightness of RR Lyr, the exposure times
were as short as possible to avoid saturation. 
Therefore, the individual exposures in the J, H and K bands were split
into twenty 0.5 sec sub-exposures that were dithered by a few pixels each, in
both X and Y directions, for a total exposure time of 10 sec in each filter. 

All images were processed using the standard procedure described in Del
Principe et al. (2005). Each sky image was obtained as the median of all the
dithered frames after cleaning off the detected sources. Median sky frames were
subtracted from each target image before flat-fielding and co-adding all the
dithered images.

The photometric analysis has been performed using the SExtractor photometric
package (Bertin \& Arnouts 1996). For each star we measured the flux contained
within a radius corresponding to $\sim$ 2 FWHM from the star center. A
comparison of the mean magnitudes measured for the 7 stars present in the same 
RR Lyr field with those reported in the 2MASS catalog indicates the absence of
any colour term in the magnitude differences. Therefore, the zero-points to be 
added to the instrumental JHK magnitudes of RR Lyr have been calculated as 
the average of the differences between the 2MASS and instrumental JHK 
magnitudes of the 7 reference stars present in the field. The r.m.s. error on 
the individual data points is $\sim$ 0.01 mag in all the considered photometric bands. 
These calibrating stars and their 2MASS catalogue data are listed in 
Table 1. 

\begin{figure*}
 \includegraphics[width=12.cm]{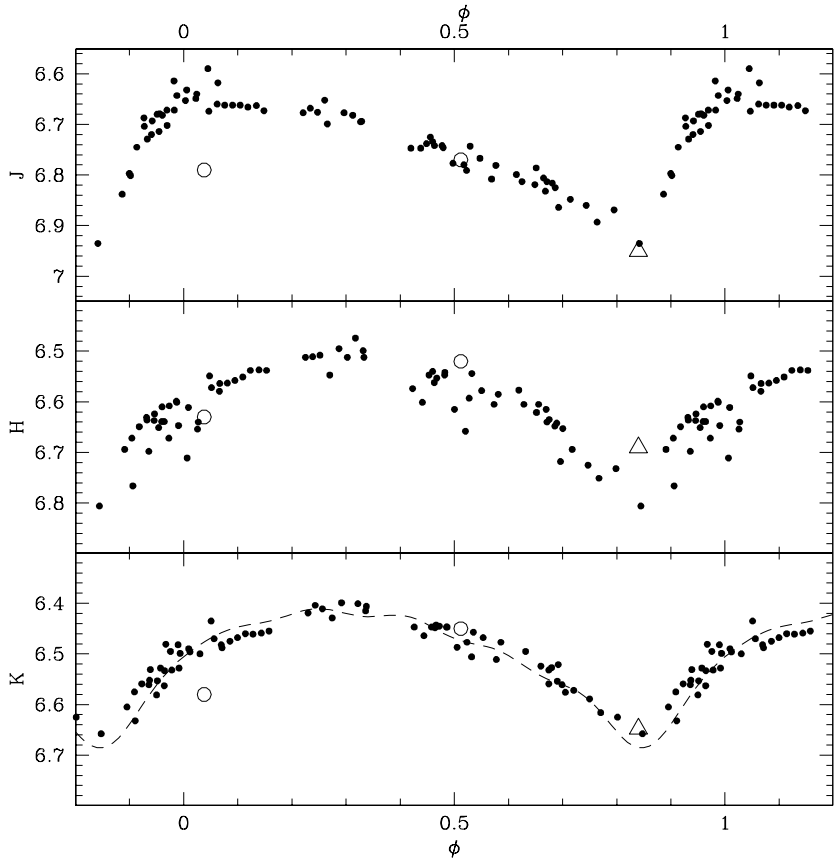}
\caption{Light curves of RR Lyr in the J, H and K bands. The dashed curve in the 
bottom panel show the template from Jones et al. (1996). The two open circles and the 
open triangle mark the FSB93 and 2MASS data points, respectively.}
\label{f:lcur}
\end{figure*}

The phases of the RR Lyr data points have been calculated using the epoch of 
maximum light $T_{0}$=2454025.494 (selected from the 
list of maxima for RR Lyr in the GEOS RR Lyr database; La Borgne et al. 2004) 
and the period P=0.5668386d (Kolenberg et al. 2006). The
corresponding JHK light curves\footnote{The calibrated JHK magnitudes for RR Lyr, along with the Heliocentric Julian 
Day (HJD) and phase of the observations, are available in electronic form at the 
CDS (http://cdsweb.u-strasbg.fr/).} are shown in Figure \ref{f:lcur}.
Note the presence of few outliers in the H light curve at phases $0.4\div0.5$ 
and $0.9\div1.0$ . 
These measures could be affected by uncertainties in the background
determination, which is more unstable in the H band than in the other passbands.

\begin{table*} 
 \centering
  \caption{The calibrating 2MASS stars. The ID and JHK magnitudes 
are taken from the 2MASS Catalogue.} 
\label{t:cal}
  \begin{tabular}{@{}lccccr@{}}
  \hline
Star ID  & RA (J2000) & Dec (J2000) & J $\pm$ err & H $\pm$ err & K $\pm$ err \\
         & deg        & deg         &             &             &             \\ 
 \hline
19252482+4247382  & 291.353429 & 42.793964  &  12.641 $\pm$ 0.021 & 12.019 $\pm$ 0.017 & 11.876 $\pm$ 0.016\\
19253010+4248407  & 291.375447 & 42.811321  &  11.977 $\pm$ 0.021 & 11.728 $\pm$ 0.018 & 11.653 $\pm$ 0.018\\
19252472+4245382  & 291.353005 & 42.760620  &  11.530 $\pm$ 0.021 & 11.280 $\pm$ 0.017 & 11.200 $\pm$ 0.012\\
19252875+4248290  & 291.369824 & 42.808071  &  12.260 $\pm$ 0.021 & 11.900 $\pm$ 0.018 & 11.807 $\pm$ 0.014\\
19252995+4247331  & 291.374810 & 42.792538  &  12.806 $\pm$ 0.021 & 12.360 $\pm$ 0.017 & 12.270 $\pm$ 0.018\\
19252353+4247403  & 291.348050 & 42.794540  &  12.395 $\pm$ 0.022 & 12.177 $\pm$ 0.017 & 12.109 $\pm$ 0.018\\
19252017+4248310  & 291.334045 & 42.808613  &  11.835 $\pm$ 0.031 & 11.279 $\pm$ 0.029 & 11.133 $\pm$ 0.025\\
 \hline
\end{tabular} 
\end{table*} 

\section{Results}
\label{res}

The mean JHK magnitudes have been derived by Fourier analysis of the light 
curves, and are 
$<$J$>$=6.74 $\pm$ 0.02, $<$H$>$=6.60 $\pm$ 0.03 and $<$K$>$=6.50 $\pm$ 0.02. 
In Table 2 the obtained mean magnitudes are compared with the 
mean values obtained by FSB93 (average of two magnitudes about 
half a period apart) and with the single epoch photometry from 2MASS taken at 
phase $\phi \sim$0.79 (i.e. close to the minimum light). 
The magnitudes measured by FSB93 have been reported in the 2MASS 
photometric system by adopting the trasformations by Carpenter et al. (2001). 
We note that FSB93's mean values agree quite well (within 0.04 mag) with 
our results. The 2MASS values reported here for completeness are obviously very 
different as they are single epoch measures corresponding to the minimum light 
phase.
Both the FSB93 and the 2MASS data points 
are shown in Fig. \ref{f:lcur}, for comparison with our light curves. 

Note that significant uncertainties can be present in the phases of the 
individual data points. 
On the assumption that the (very accurate) value of the period holds for all 
data sets, errors in the phase can arise from uncertainties in 
the definition of the JD of maximum light, and are minimised by the use of 
maxima as close in time as possible to the data. 
For the phases of FSB93
data we used an epoch of maximum light T$_0$=2445556.4304 (Huebscher \& Mundry 1984). 
In this case we estimate that the uncertainty in the phase is less than 0.02. 
For the 2MASS data there are no estimates of maxima very close to the epoch 
of the observation and the O-C (from visual estimates) of the JD max 
at that time show strong variations. By averaging the results obtained from 
JD max estimates in the time range of approximately $\pm$ one year of the 
2MASS observation we obtain $\phi$=0.79$\pm$0.05.

In the K panel of Fig. \ref{f:lcur} we show also the template K light curve 
obtained by Jones et al. (1996) from the average behaviour of RR Lyrae variables. 
The selected template corresponds to a blue light curve amplitude of about 1 mag, 
which is appropriate for RR Lyr (Hardie 1955). The template, shifted to reproduce $<$K$>$=6.50, 
matches our observed data points within $\pm$0.05 mag at all phases and the 
overall agreement is quite good. 
 
\begin{table}
\label{t:comp}
 \centering
  \caption{Comparison between the JHK magnitudes of RR Lyr measured in this work 
  and in the literature.}
  \begin{tabular}{@{}lccr@{}}
  \hline
  &This work & FSB93 & 2MASS\\
  & mean mag. & mean mag. &   $\phi \sim$~0.79 \\
 \hline
$J$  & 6.74 & 6.78 & 6.95\\
$H$  & 6.60 & 6.57 & 6.69\\
$K$  & 6.50 & 6.51 & 6.65\\
 \hline
\end{tabular} 
\end{table} 

\subsection{Blazhko effect}

Our observations cover a range of 48 days, which is more than the Blazhko 
period of 38.4 d (Kolemberg et al. 2006). 
This explains why the scatter in the light curves is slightly larger 
than expected from the photometric errors on the individual data points. 
However, we could not identify unambiguously light curves corresponding to 
different Blazhko phases because the nightly coverage of the pulsation cycle was 
not sufficient. 

A study of Blazhko stars by Jurcsik et al. (2002) has shown that RR Lyr displays 
distorted light curves at all phases of the 4-yr cycle except at low modulation 
amplitude phase, when the shape of the light curves is similar to that of 
non-Blazhko stars. 
Our observations are taken during a period 
of maximum modulation amplitude, when the light curves are expected to be 
significantly distorted with respect to the ``normal'' behaviour.  
However, according to a study of Blazhko stars in the globular cluster M3
(Cacciari et al. 2005), the Blazhko modulation does not seem to affect 
significantly the {\em mean} magnitude.
Therefore, we conclude that the values derived above 
are a reliable representation of the mean J, H and K magnitudes 
of RR Lyr.   

\subsection{Reddening}
\label{redd}

Visual-infrared combined colours present a larger dependence on reddening than optical ones 
($E(V-K)/E(B-V)=2.63$; Cardelli et al. 1989). 
For this reason, the mean $<V>-<K>$ 
colour can be used to constrain the reddening coefficient E(B$-$V) of RR Lyr.  
To this purpose, we used the period-colour-metallicity  
relation obtained from Catelan et al. (2004) 
period-luminosity-metallicity relations in V and K\footnote{We 
converted the metal mass fraction Z in metallicity [Fe/H] using eq. 9 and 10
of Catelan et al. (2004).} (see their eqn. 6 and 7).
After correcting Catelan et al.'s K photometry to the 
2MASS system using Carpenter (2001; eq. A.1), this relation predicts:

\begin{equation}
\label{eq1}
<V_{0}>-<K_{0}>=2.929+2.353~log~P  \\
\end{equation}
~~~~~~~~~~~~~~~~~~~~~~~~~$+0.108~(log~Z)^2+0.707~log~Z$

Using $[Fe/H]=-1.39$ and $[\alpha/Fe]=0.31$ (Clementini et al. 1995) and the 
value of the period as in Sect. \ref{reduct}, we obtain $<V_{0}>-<K_{0}>$=1.20. 
The observed mean V magnitude has been obtained by Fourier analysis of the 
light curve provided by Smith et al. (2003), and is $<V>=7.75\pm0.05$.
The difference between the observed and predicted mean colours allows us to 
estimate the effect of reddening, and to derive $E(V-K)=0.05\pm$0.03 and 
hence $E(B-V)=0.02\pm0.03$. 
The uncertainty on the derived reddening coefficient has been estimeted by
using the standard propagation of the errors in the coefficients given by 
Catelan et al. (2004), in the mean V and K magnitude and the metallicity.
Previous reddening determinations estimated a value of $E(B-V)=0.03\pm0.02$
(Burstein \& Heiles 1978; Blanco 1992) and
$E(B-V)=0.06\pm0.03$ (Fernley et al. 1998). 
On the other hand, Benedict et al. (2002) assumed a reddening coefficient 
corresponding to A$_V$=0.07$\pm$0.03 as the mean from five field 
stars close to RR Lyr. This value, however, could be as high as 
A$_V$=0.11$\pm$0.10 at the location of RR Lyr if a variation across the field 
was present and a two-dimensional linear interpolation in A$_V$ between 
the four nearest astrometric reference stars was performed. 
All of these values are compatible within the errors with the estimate we 
have derived. In the following we adopt 
$E(B-V)=0.02\pm0.03$ as derived above, 
and hence A$_K$=0.006 and A$_V$=0.054 assuming $A_{K}/E(B-V)=0.35$ and 
$A_{V}/E(B-V)=3.1$ (Cardelli et al. 1989).  

\subsection{Mean Temperature and Luminosity}
\label{TL}

The $V-K$ colour is one of the best temperature indicator. 
Indeed, it has a larger sensitivity to the photospheric temperature 
and a smaller dependence on metallicity with respect to optical colours 
(Alonso et al. 1999).

To estimate the mean temperature of RR Lyr we used the color-temperature 
transformations by Castelli (1999) and Montegriffo et al. (1998). 
For this purpose, we corrected the measured mean 
colour $<V>-<K>$=1.25 for the reddening $E(V-K)=0.05$ and applied a further 
correction of $-$0.02 mag to account for non-static atmospheric effects 
(Bono et al. 1995a). The resulting intrinsic static colour $<V>-<K>$=1.18 
in the 2MASS system assumes the values of 1.14 in the Bessell \& Brett (1988) 
system adopted by Castelli (1999), and 1.15 in the system adopted by 
Montegriffo et al. (1988). Therefore the corresponding 
temperatures are T$_{eff}$=6423 K, 6253 K and 6313 K using  
Castelli (1999) calibration, and Montegriffo et al. (1998) empirical and 
theoretical calibrations, respectively (see Table 4 in Cacciari et al.. 2005). 
The mean of these values is $<T_{eff}>=6330\pm50~^o K$. 

The luminosity of RR Lyr can be estimated once a distance modulus is known.
Considering the two determinations of trigonometric parallax quoted 
in Sect. 1, we adopt the value by Benedict et al. (2002) for its much smaller 
error, and hence $(m-M)_0=7.09\pm0.11$ and $M_K=-0.60\pm0.03$.
Using BC$_K$=1.15 from Montegriffo et al. (1998)
and $M^{bol}_{\odot}=4.75$ (the IAU convention) we then derive the 
luminosity of RR Lyr $log~L/L_{\odot}=1.68\pm0.05$.
This estimate is in good agreement with that reported by Bono et al. 
(2003) using a theoretical period-K luminosity relation 
($log~L/L_{\odot}=1.689\pm0.030$). 

\begin{figure}
 \includegraphics[width=8.7cm]{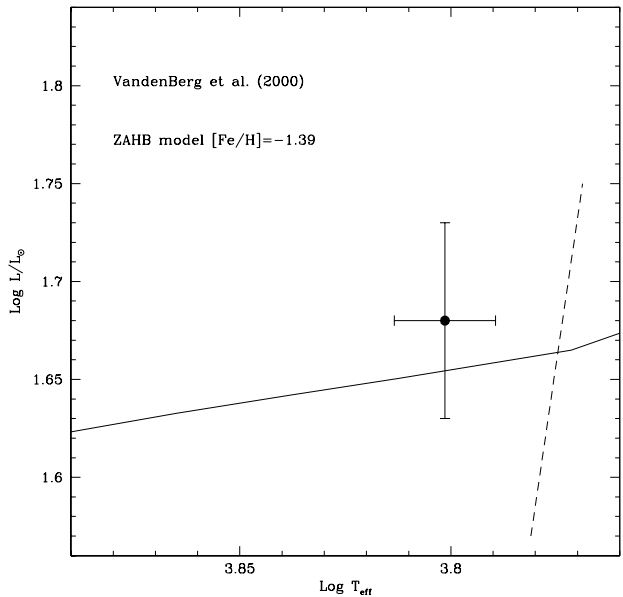}
\caption{Location of RR Lyr in the Hertzprung-Russel diagram. 
The theoretical ZAHB of VandenBerg et al. (2000) and the red 
edge of the instability strip by Bono et al. (1995b) are overplotted.}
\label{hr}
\end{figure}

In Fig. \ref{hr} we show the location of RR Lyr in the theoretical 
Hertzprung-Russel diagram. For comparison, the theoretical Zero Age Horizontal 
Branch (ZAHB) interpolated to [Fe/H]=$-$1.39 from the set of ZAHB models by 
VandenBerg et al. (2000) is overplotted along with the red edge of the 
instability strip predicted by Bono et al. (1995b).
As can be noted, RR Lyr appears to be located close to  
the theoretical ZAHB and not far from the red edge of the instability strip.
Given the uncertainties both on the data and on the theoretical models, this 
indicates that the star has not yet significantly evolved from the phase of 
core helium burning. 
A further confirmation of this result is provided by the location of RR Lyr in
the B amplitude-period relation (see Cacciari et al. 2005). Indeed, in this
diagram the maximum B amplitude of RR Lyr places it close to the main locus of 
Oostheroff I variables and not above it as would be the case for an evolved 
star.

The comparison with the ZAHB model yields an 
estimate of the mass of RR Lyr of $M=0.63\pm0.01~M_{\odot}$. 

\section{The zero-point of the distance scale} 

In Sollima et al. (2006) RR Lyr was used to set the zero-point of the period - 
infrared luminosity relation derived from the analysis of 544 RR Lyrae variables 
in 16 globular clusters. To that purpose, the M$_K$ magnitude of RR Lyr was 
estimated to be $-$0.57 mag based on the data and knowledge then available. 
We can now provide a more reliable and accurate value, based on the present 
infrared light curves and updated values of reddening, namely $M_K=-0.60$ mag. 
This value is 0.03 mag brighter than the previous one, and the eq.s (4) and 
(5) in Sollima et al. (2006) become: 
\begin{equation}
\label{eq2}
 M_{K} =  - 2.38(\pm 0.04)~log~P_{F} + 0.08(\pm 0.11)~[Fe/H]_{CG}\\
\end{equation}
~~~~~~~~ $- 1.07(\pm 0.11) $

\smallskip
\noindent where $P_{F}$ is the fundamental period and $[Fe/H]_{CG}$ the
metallicity in the Carretta \& Gratton (1997) metallicity scale. 

The application of these log P-M$_K$-[Fe/H] relations to the RR Lyrae 
stars in the globular cluster Reticulum in the LMC leads to a distance modulus 
$(m-M)_{0}=18.48\pm0.11$, using the K data by Dall'Ora et al. (2004) and 
$[Fe/H]=-1.71$ and $E(B-V)=0.03$ (Suntzeff et al. 1992).
On the other hand, the absolute V magnitude of RR Lyr derived here 
(M$_V$=0.61$\pm$0.12) can be also used to set the zero point of the 
period - V luminosity relation.
Assuming $\delta V/\delta [Fe/H]=0.214$ (Clementini et al. 2003) we 
derive M$_V$=0.54 at $[Fe/H]=-1.71$ (as appropriate for Reticulum). 
This value, together with the mean 
dereddened V magnitude $<V_{0}>$=18.98 for these variable stars (from Walker 1992)
implies a distance to the Reticulum of $(m-M)_{0}=18.44\pm0.14$.
The agreement of the V and K results 
make us confident that we have provided the most reliable and accurate 
JHK data presently available for RR Lyr. 

As an interesting application of this infrared relation, we can estimate the distance 
to the set of RR Lyrae 
stars observed by Borissova et al. (2004) in the inner regions of the LMC 
(see Sollima et al. 2006), which turns out to be $(m-M)_{0}=18.56\pm0.13$.
Similarly, we can estimate the distance to the bar of the LMC, where about 
100 RR Lyrae stars with an average metallicity of 
[Fe/H]=$-$1.5 (Clementini et al. 2003) have a mean 
dereddened V magnitude $<V_{0}>$=19.064. 
Using the absolute V magnitude of RR Lyr derived above we obtain 
$(m-M)_{0}=18.48\pm0.13$. 
This result is also in 
good agreement with the most recent studies of Cepheid variables 
(Persson et al. 2004; Gieren et al. 2005; van Leeuwen et al. 2007; 
Fouqu\'e et al. 2007).
The difference between the distance to the bar (from the V relation) and 
to the inner field (from the K relation), which is larger than the 
difference found for the Reticulum, may indicate
that the two fields are not exactly at the same distance.

\section{Conclusions}

In this paper we provide the first light curves of the variable star RR Lyr 
in the infrared JHK passbands, yielding to the mean magnitudes  
$<$J$>$=6.74 $\pm$ 0.02, 
$<$H$>$=6.60 $\pm$ 0.03 and $<$K$>$=6.50$\pm$ 0.02. 

The mean temperature and luminosity of RR Lyr have been derived. 
These quantities have been used to estimate the mass of this star and 
its evolutionary stage by comparison with theoretical ZAHB models of 
appropriate metallicity. This comparison indicates that RR Lyr is located on 
or very near the ZAHB locus, and so it is still burning helium in its core. 
 
The excellent quality of the present infrared data allowed us to use RR Lyr 
as a reliable and accurate infrared standard candle to set the zero point of 
the distance scale. In particular, the mean $M_K=-0.60$ magnitude was used to 
calibrate the period - K luminosity - metallicity relation for RR Lyrae stars. 

The most important aspects of these result are: 

i) With the present calibration we derive a distance modulus to the globular 
cluster Reticulum of the LMC of $(m-M)_{0}=18.48\pm0.11$.
This value, obtained from the $<K>$ of RR Lyr, is comparable to the value 
obtained from the $<V>$ ($(m-M)_{0}=18.44\pm0.14$) within the errors. 
This indicates that the mean V and K magnitudes of RR Lyr are 
accurate and self-consistent.  

ii) The estimated distances to the bar and an inner field of the LMC are also 
in good agreement with the most 
recent studies of Cepheid variables (Persson et al. 2004; Gieren et al. 2005; 
van Leeuwen et al. 2007; Fouqu\'e et al. 2007) although there is some indication 
that the inner field is slightly more distant.   

\section*{acknowledgements} 
We warmly thank Paolo Montegriffo for assistance during catalogs cross-correlation and 
Mauro Dolci for his technical support for the observations and data reduction procedures. 
We are grateful to the anonymous referee for his comments and suggestions 
that have contributed to improve the paper. 
This research was supported by contract ASI-INAF I/023/05/0 and PRIN-INAF 2006.

\label{lastpage}

\end{document}